\documentclass[aps,12pt,final,showpacs,showkeys,onecolumn,superscriptaddress]{revtex4}%
\usepackage{amsfonts}
\usepackage{amsmath}
\usepackage{amssymb}
\usepackage{graphicx}%
\providecommand{\U}[1]{\protect\rule{.1in}{.1in}}
\begin{document}

\title{Electron beam - plasma interaction
in a dusty plasma \\with excess suprathermal electrons }

\author{A. Danehkar}
\affiliation{Department of Physics and Astronomy, Macquarie
University, Sydney, NSW 2109,
  Australia}
\author{N. S. Saini}
\affiliation{Department of Physics, Guru Nanak Dev University,
Amritsar-143005, India}

\author{M. A. Hellberg}{
\affiliation{School of Physics, University of KwaZulu-Natal, Durban
4000, South Africa}

\author{I. Kourakis}{
\affiliation{Department of Physics and Astronomy, Queen's University
Belfast, BT7 1NN, UK}


\begin{abstract}
The existence of large-amplitude electron-acoustic solitary
structures is investigated in an unmagnetized and collisionless
two-temperature dusty plasma penetrated by an electron beam. A
nonlinear pseudopotential technique is used to investigate the
occurrence of stationary-profile solitary waves, and their
parametric dependence on the electron beam and dust perturbation is
discussed.
\end{abstract}

\maketitle

%%%%%%%%%%%%%%%%%%%%%%%%%%%%%%%%%%%%%%%%%%%%
%% MAINMATTER
%%%%%%%%%%%%%%%%%%%%%%%%%%%%%%%%%%%%%%%%%%%%

%\section{<A section>}

%
% Abstract text.
%

%\textbf{1. Introduction.}
We have previously studied electron-acoustic solitary waves in the
presence of a suprathermal electron component. \cite{Danehkar2011}
Our aim here is to investigate the effect of beam electrons and dust
on the electrostatic solitary structures.

%\textbf{2. Model Equations.}
We consider a plasma consisting of cold inertial drifting electrons
(the beam), cold inertial background electrons, hot suprathermal
electrons modeled by a kappa-distribution, stationary ions, and
stationary dust (of either positive or negative charge). The
dynamics of the cold inertial background electrons and the beam
electrons are governed by the following
normalized one-dimensional equations:%
\begin{equation}
\begin{array}{cc}
\dfrac{\partial n}{\partial t}+\dfrac{\partial (nu)}{\partial x}=0, & \dfrac{%
\partial u}{\partial t}+u\dfrac{\partial u}{\partial x}=\dfrac{\partial \phi
}{\partial x},%
\end{array}
\label{eq1}
\end{equation}%
\begin{equation}
\begin{array}{cc}
\dfrac{\partial n_{b}}{\partial t}+\dfrac{\partial (n_{b}u_{b})}{\partial x}%
=0, & \dfrac{\partial u_{b}}{\partial t}+u_{b}\dfrac{\partial u_{b}}{%
\partial x}=\dfrac{\partial \phi }{\partial x},%
\end{array}
\label{eq2}
\end{equation}%
\begin{equation}
\frac{\partial ^{2}\phi }{\partial x^{2}}=-\left( \eta +s\delta
\right)
+n+\beta n_{b}+(\eta +s\delta -1-\beta )\left( 1-\frac{\phi }{[\kappa -%
\tfrac{3}{2}]}\right) ^{-\kappa +1/2}.  \label{eq3}
\end{equation}%
Here, $n$ and $n_{b}$ denote the fluid density variables of the cool
electrons and the beam electrons normalized with respect to $n_{c,0}$ and $%
n_{b,0}$. The velocities $u$ and $u_{b}$, and the equilibrium beam speed $%
U_{0}=u_{b,0}/c_{th}$ are scaled by the hot electron thermal speed
$c_{th}=\left(
k_{B}T_{h}/m_{e}\right) ^{1/2}$, and the wave potential $\phi $ by $%
k_{B}T_{h}/e$. Time and space are scaled by the plasma period
$\omega _{pc}^{-1}=\left( n_{c,0}e^{2}/\varepsilon _{0}m_{e}\right)
^{-1/2}$and the characteristic length $\lambda _{0}=\left(
\varepsilon _{0}k_{B}T_{h}/n_{c,0}e^{2}\right) ^{1/2}$,
respectively. We define the hot-to-cold electron charge density
ratio $\alpha =n_{h,0}/n_{c,0}$, the beam-to-cold electron charge
density ratio $\beta =n_{b,0}/n_{c,0}$, the ion-to-cold electron
charge density ratio $\eta =Z_{i}n_{i,0}/n_{c,0}$, and the
dust-to-cold electron charge density ratio $\delta
=Z_{d}n_{d,0}/n_{c,0}$. Here, suprathermality is denoted by the
spectral index $\kappa $, and $s=\pm 1$ is the sign of the dust
charge for positive or negative dust grains. At equilibrium, the
plasma is quasi-neutral, so $\eta +s\delta =1+\alpha +\beta$.

% figure1[htb]
\begin{figure}[ptb]
\centering
\includegraphics[width=6.4in]{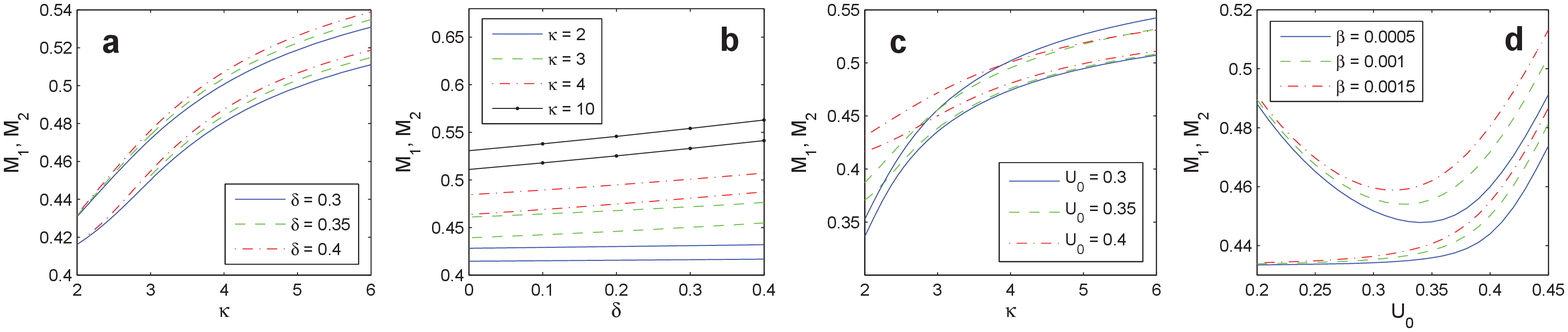}
\caption{Soliton existence region ($M_1 < M < M_2$): (a) versus
$\kappa$ for different $\delta$ values; (b) versus $\delta$ for
different $\kappa$ values; (c) versus $\kappa$ for different $U_0$
values; (d) versus $U_0$ for different $\beta$ values. The remaining
values are $\kappa=4.5$, $\delta=0.3$, $s=-1$, $\beta=0.001$, $U_0
=0.4$ and $\eta=4.5$, unless values are given.}%
\label{figure1}%
\end{figure}

%\textbf{4. Nonlinear Analysis.}
Anticipating constant profile solutions of Eqs.
(\ref{eq1})--(\ref{eq3}) in a stationary frame traveling at a
constant normalized velocity $M$, implying the transformation $\xi
=x-Mt$, we obtain $n=(1+2\phi/M^{2})^{-1/2}$ and $n_{b}=[ 1+2\phi /
(M-U_{0})^{2}]^{-1/2}$. Substituting in Poisson's equation and
integrating yields a pseudo-energy balance equation $\frac{1}{2%
}\left( d\phi /d\xi \right) ^{2}+\Psi (\phi )=0$, where the Sagdeev
pseudopotential $\Psi (\phi )$ reads%
\begin{eqnarray}
&& \Psi (\phi ) = \left( \eta +s\delta \right) \phi +M^{2}\left(
1-[1+2\phi /M^{2}]^{\frac{1}{2}}\right) +\beta (M-U_{0\text{ }})^{2}
\times \notag
\\
&& \Big(1-[1+\left. 2\phi /(M-U_{0\text{
}})^{2}]^{\frac{1}{2}}\right) +(\eta +s\delta
-1-\beta )\left( 1-[1-\phi /(\kappa -\tfrac{3}{2})]^{-\kappa +\frac{3}{2}%
}\right) .
\end{eqnarray}%
Reality of the density variable implies two limits on the
electrostatic
potential $\phi _{\max }=-M^{2}/2$ and $-(M-U_{0\text{ }})^{2}/2$ for $%
U_{0}<0$ and $U_{0}>0$, respectively. In order for solitary waves to
exist, two constraints must be satisfied, i.e., $F_{1}(M)=-\Psi
^{\prime \prime }(\phi )|_{\phi =0}>0$ and $F_{2}(M)=\Psi (\phi
)|_{\phi =\phi _{\max }}>0$, which yield the solutions for the lower
and upper limit in $M$.
%The soliton existence domains obtained are depicted in Figure
%\ref{figure1}.

As shown in Figure \ref{figure1}, the existence domain for solitons
becomes narrower with increasing suprathermal excess (decreasing
$\kappa$), increasing equilibrium beam speed, and decreasing beam
density. Dust charge density shows little effect on the width of the
existence domain, but for quasi-Maxwellian electrons, it weakly
increases the typical values of $M$.  It was  found that both
increasing $\kappa$ and increasing negative dust charge density
significantly reduce soliton amplitude at fixed $M$ (not shown
here).
%It is thus confirmed that both dust and an electron beam have
%important effects on soliton behaviour.

\section{Acknowledgments}
%{\footnotesize \textbf{Acknowledgment.}
%\textbf{Acknowledgments.}
AD, NSS and IK thank the Max-Planck Institute for Extraterrestrial
Physics for their support. IK acknowledges support from UK EPSRC via
S\&I grant EP/D06337X/1.
%AD, NSS and IK acknowledge financial support from the Max-Planck
%Institute for Extraterrestrial Physics. IK acknowledges support from
%UK EPSRC via S\&I grant EP/D06337X/1.

\end{document}